\pdfoutput=1

\documentclass[11pt]{article}

\usepackage[final]{acl}

\usepackage{times}
\usepackage{latexsym}
\usepackage{xurl}
\usepackage[T1]{fontenc}

\usepackage[utf8]{inputenc}

\usepackage{microtype}

\usepackage{inconsolata}

\usepackage{graphicx}
\usepackage{multirow} 
\usepackage{orcidlink}
\usepackage{booktabs}

\newcommand\blfootnote[1]{
    \begingroup
    \renewcommand\thefootnote{}\footnote{#1}
    \addtocounter{footnote}{-1}
    \endgroup
}

\title{ElliottAgents: A Natural Language-Driven Multi-Agent System for Stock Market Analysis and Prediction}

\author{Jarosław A. Chudziak \orcidlink{0000-0003-4534-8652} \\
  Warsaw University of Technology \\
  Institute of Computer Science \\
  Warsaw, Poland \\
  0000-0003-4534-8652 \\
  \texttt{jaroslaw.chudziak@pw.edu.pl} \\\And
  Michał Wawer \orcidlink{0009-0004-2717-1616}\\
  Warsaw University of Technology \\
  Institute of Control and Computation Engineering \\
  Warsaw, Poland \\
  0009-0004-2717-1616 \\
  \texttt{michal.wawer.stud@pw.edu.pl} \\
}

\begin{document}
\maketitle
\begin{abstract}
This paper presents ElliottAgents, a multi-agent system leveraging natural language processing (NLP) and large language models (LLMs) to analyze complex stock market data. The system combines AI-driven analysis with the Elliott Wave Principle to generate human-comprehensible predictions and explanations.
A key feature is the natural language dialogue between agents, enabling collaborative analysis refinement. The LLM-enhanced architecture facilitates advanced language understanding, reasoning, and autonomous decision-making. Experiments demonstrate the system's effectiveness in pattern recognition and generating natural language descriptions of market trends.
ElliottAgents contributes to NLP applications in specialized domains, showcasing how AI-driven dialogue systems can enhance collaborative analysis in data-intensive fields. This research bridges the gap between complex financial data and human understanding, addressing the need for interpretable and adaptive prediction systems in finance.
\end{abstract}

\blfootnote{This is the accepted version of the paper presented at the \textbf{38th Pacific Asia Conference on Language, Information and Computation}, Tokyo, Japan. Available at: \url{https://aclanthology.org/2024.paclic-1.91/}}

\section{Introduction}

The integration of LLMs into multi-agent systems has opened new frontiers in AI, particularly in the domain of financial analysis \citep{zhao_survey_llm_agents_2023, weng_llm_agents_2024}. Stock market prediction, a field characterized by its complexity and dynamism, has long challenged traditional AI-based methods. These approaches often falter in processing vast datasets and adapting to rapid market changes \citep{gamil_mas_fuzzy_logic_stock_market_2007, luo_mas_decision_support_stock_market_2002}.

This paper presents ElliottAgents, an multi-agent system that harnesses the power of NLP \citep{lane_nlp_in_action_2019} and LLMs to analyze stock market data. Our approach combines AI-driven analysis with the Elliott Wave Principle (EWP), a established method of technical analysis \citep{frost_ewp_2001}. The core innovation lies in the system's ability to facilitate natural language dialogue between agents, enabling them to collaboratively interpret market patterns and refine their analyses.

Our research addresses the following question: How can we effectively integrate natural language processing methods and multi-agent architectures to produce reliable and human-comprehensible stock market analyses and predictions? Through experimental validation, we demonstrate that our approach not only enhances pattern recognition accuracy but also generates detailed, easily interpretable market trend descriptions and forecasts.

Our system contributes to the field of NLP applications in specialized domains. It showcases the potential of AI-driven dialogue systems in enhancing collaborative analysis within data-intensive fields. By filling the gap between complex financial data and human understanding, ElliottAgents represents a step forward in creating more interpretable and adaptive prediction systems in finance.

\section{Foundations of Stock Market Forecasting}

\subsection{The Evolution of Stock Market Analysis}

Stock market analysis has progressed from manual techniques to AI-driven approaches over the past century. Traditional methods like fundamental analysis and technical analysis \citep{murphy_technical_analysis_1999} have been augmented by computational models since the 1960s. The development of financial software has seen several paradigm shifts: from simple automation of existing techniques to the creation of complex algorithmic trading systems \citep{tirea_mas_ewp_2012}. Recent years have witnessed the integration of machine learning and natural language processing in financial analysis. 
Our proposed ElliottAgents system addresses several limitations in current market analysis approaches. The system's distributed nature enables parallel processing of market data, allowing for real-time analysis across numerous assets and timeframes simultaneously. In comparison to other approaches, ElliottAgents offers a more interpretable framework by integrating the structured EWP approach. Our recommendations are based on theory that has been used for years in contrast to the “blackbox” nature of other AI-based systems. This integration potentially provides a longer-term perspective and strategies more aligned with established market behavior patterns. The system's ensemble of specialized agents, each focusing on different aspects of EWP analysis, aims to provide a more holistic market view compared to purely data-driven ensembles.

\subsection{Elliott Wave Principle}

\begin{figure}[!t]
    \centering
    \includegraphics[width=\columnwidth]{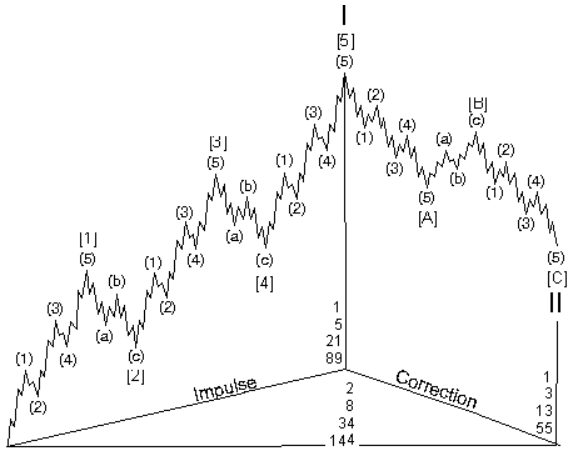}
    \caption{The fractal character of Elliott wave pattern \citep{frost_ewp_2001}}
    \label{fig:elliott_example}
\end{figure}

The Elliott Wave Principle (EWP), developed by Ralph Nelson Elliott, is a technical analysis method based on the premise that market prices move in recognizable patterns driven by collective investor psychology \citep{frost_ewp_2001, murphy_technical_analysis_1999}. This principle posits that market behavior alternates between phases of optimism and pessimism, creating predictable waves in price movements. Elliott identified thirteen recurring patterns, or "waves," which can be classified into two main types: impulsive and corrective waves. As presented in Fig. \ref{fig:elliott_example}, impulsive waves are the driving force behind market trends and consist of five sub-waves, while corrective waves, comprising three sub-waves, counterbalance the trend.

\begin{figure}
    \centering
    \includegraphics[width=\columnwidth]{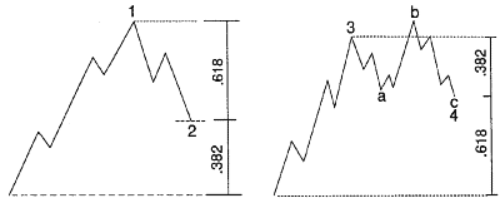}
    \caption{Fibonacci retracements in corrective waves \citep{frost_ewp_2001}}
    \label{fig:fibo_retracements}
\end{figure}

The Fibonacci sequence plays a crucial role in the EWP, providing a mathematical framework for wave relationships \citep{boroden_fibo_trading_2008}. Elliott observed that waves often align with Fibonacci ratios, particularly the Golden Ratio (approximately 1.618). These ratios govern the relative lengths and amplitudes of waves, with Wave 3 in an impulsive sequence typically being 1.618 times the length of Wave 1. Corrective waves often retrace Fibonacci percentages (38.2\%, 50\%, 61.8\%) of the previous impulsive wave as presented in Fig. \ref{fig:fibo_retracements}.

The fractal nature of Elliott waves \citep{vantuch_ewp_forecasting_2016} allows for application across various time frames, from short-term movements to long-term trends. This characteristic, combined with Fibonacci relationships, creates a cohesive structure throughout market cycles . While the EWP does not offer certainty, it provides a framework for assessing probabilities of different market scenarios, aiding traders in understanding market context and predicting potential future paths.

\subsection{Large Language Models}

Large language models represent a significant advancement in the field of AI and NLP. These models are designed to understand, generate, and interact with human language in a way that is increasingly indistinguishable from human performance \citep{naveed_overview_llms_2024, khan_raiaan_llms_review_2024}.  Examples of LLMs include OpenAI's GPT-4 \citep{open_ai_gpt4_2023}, Google's Gemini, and the LLAMA series. The advancement of NLP is intrinsically tied to the progress of LLMs. These models, which lie at the forefront of AI, are capable of understanding, generating, and interacting with human language in a way that is increasingly indistinguishable from human performance  \citep{bouchard_enhancing_llm_2024}. They have been trained on vast amounts of text data and leverage sophisticated architectures to perform a wide range of language-related tasks, from translation and summarization to question answering and creative writing. 

The core principle behind LLMs is the use of neural networks (NN) \citep{szydlowski_chudziak_nn}, specifically a type of network known as the transformer. Transformers have revolutionized the way models process sequential data. Unlike traditional recurrent neural networks (RNNs) and long short-term memory networks (LSTMs), transformers can handle long-range dependencies more effectively, making them ideal for language tasks \citep{amaratunga_understanding_llms_2023}. Transformers utilize a mechanism called self-attention, which allows the model to weigh the importance of different words in a sentence when making predictions. This is crucial for understanding context, as the meaning of a word often depends on the surrounding words.  

While LLMs have shown potential in various NLP tasks, their application in time series prediction, particularly in finance, is an area of ongoing exploration \citep{tang_time_series_forecasting_llms_2024}. One challenge is the need for LLMs to understand the temporal order of data points \citep{chudziak_nn_market_predictions_2023}, which is crucial for accurate forecasting \citep{chudziak_cinkusz_llm}. Techniques like positional encoding are used to address this limitation \citep{tan_are_llm_useful_in_time_series_2024}, but further research is needed to fully leverage the capabilities of LLMs in capturing the dynamics of financial time series. The use of agents may be a factor that will greatly improve the results of time series prediction by distributing tasks among agents, enabling a more robust analysis of complex big sets of data.

\section{Multi-Agent System Architecture}

\subsection{System Architecture}

Multi-agent systems have long been a powerful tool in modeling complex systems, where multiple autonomous entities, known as agents, interact within an environment to achieve individual or collective goals \citep{guo_llms_mas_survey_2024}. Historically, these systems were built using various methodologies, including rule-based systems, symbolic equations, stochastic modeling, and early forms of machine learning. 

However, these early approaches faced significant limitations. Agents were typically limited in their adaptability and often failed to respond effectively to dynamic, changing environments. Their interactions were straightforward, lacking the depth needed to mimic real-world complexities and making suboptimal decisions based on limited information and computational power.

The integration of LLMs, such as GPT-4 \citep{open_ai_gpt4_2023}, has significantly transformed multi-agent systems, bringing advanced natural language understanding, reasoning, and decision-making capabilities to agents. LLMs enable agents to operate more autonomously, adapting to new situations without requiring explicit instructions. These agents can now exhibit goal-directed behaviors, making proactive decisions to achieve long-term objectives, enhancing their autonomy and proactiveness \citep{guo_llms_mas_survey_2024, cinkusz_mas_project_management_2024}.
Agents can also dynamically perceive and respond to changes in their environment, learning from their experiences to improve future responses \citep{zhao_survey_llm_agents_2023, yao_react_2023}.

\begin{figure}[!t]
    \centering
    \includegraphics[width=\columnwidth]{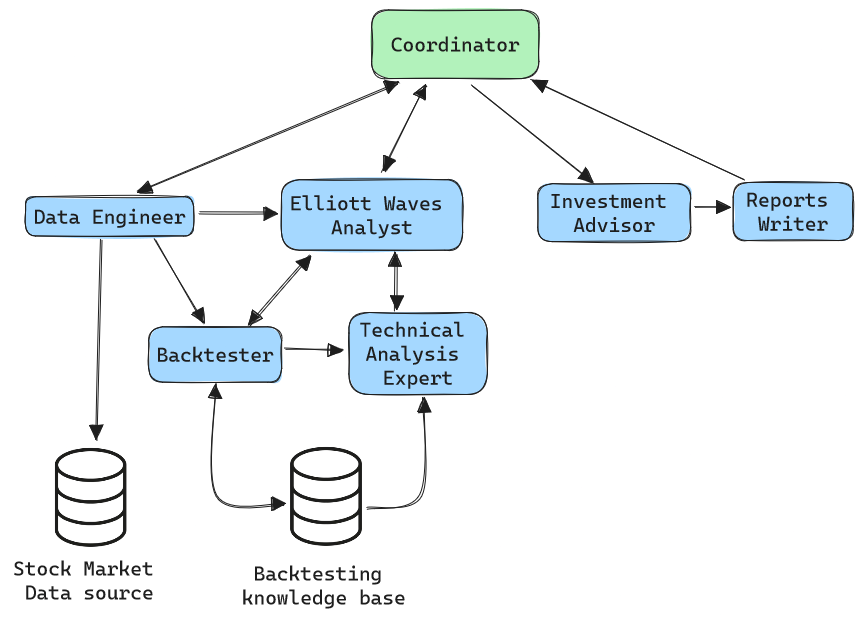}
    \caption{Data flow between agents.}
    \label{fig:agents_hierarchical}
\end{figure}

The agents collaborate, performing sequential and hierarchical tasks that culminate in a comprehensive analysis as shown on Fig. \ref{fig:agents_hierarchical}. Some agents utilize advanced tools, which were described in section "3.2 Agents Customization". 

\begin{itemize}
\item \textbf{Data Engineer}: The primary goal of this agent is to prepare the necessary data, that other agents will use for their analyses. This agent uses a dedicated tool that requires the name of the company, the timeframe and the interval for which the data is to be prepared, this information is provided by the user.
\item \textbf{Elliott Waves Analyst}: Main task of this agent is to perform detailed Elliott waves analysis on historical stock data. To do this, we create a special tool that finds all possible impulsive and corrective wave patterns in the data. Results of this tool are used to plot charts with overlaid waves at appropriate points. 
\item \textbf{Backtester}: This agent uses DRL to test and validate the predictions made by Eliott waves analyst. This process allows us to identify patterns that have worked in the past on the asset that is currently analyzed, thereby we are increasing the chances of a successful analysis. The agent uses database to retain information across past tests and can delegate tasks to other agents if necessary. 
\item \textbf{Technical Analysis Expert}: This agent's goal is to interpret the waves patterns with results of backtesting and choose the most likely pattern to occur in the current market state.
\item \textbf{Investment Advisor}: Is responsible for synthesizing various analyses into comprehensive investment strategy. This agent uses data retrieved from RAG tool and leverages the output provided by other agents. The result of his actions is an accurate investment plan, including price levels, dates, buy or sell signal and backup plans when the future stock price does not follow the predicted trends. 
\item \textbf{Reports Writer}: Summarises the activities of all agents, creating a clear, easy-to-understand report for the end user. The final output is a comprehensive report that provides clear investment strategies, including when and where to buy or sell stocks, ensuring that the recommendations are up-to-date and relevant for today's market conditions.
\end{itemize}

\subsection{Agents Customization}

\begin{figure}[!t]
    \centering
    \includegraphics[width=\columnwidth]{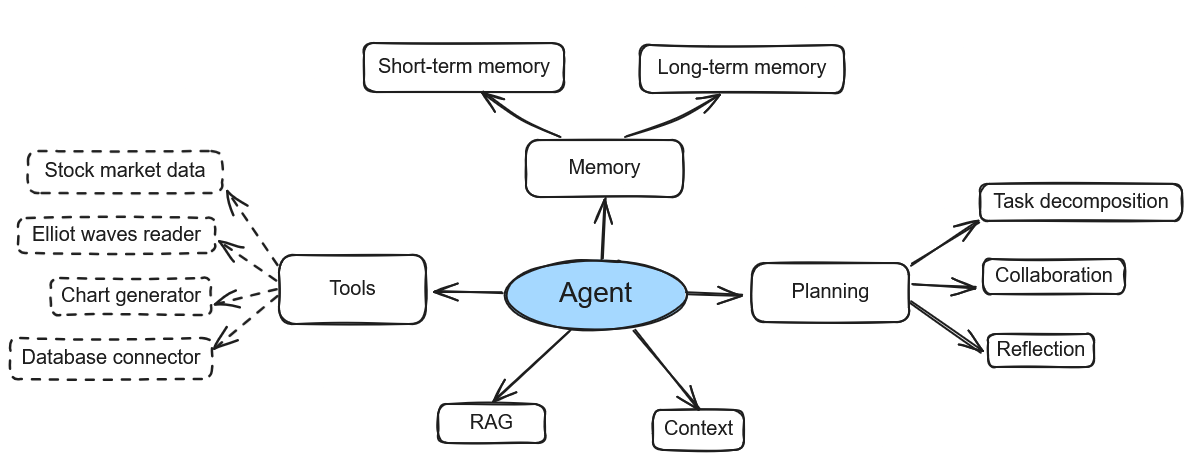}
    \caption{Overview of a LLM autonomous agent, based on \citep{weng_llm_agents_2024}.}
    \label{fig:agent}
\end{figure}

As presented on Fig. \ref{fig:agent} agent is build using  different components, most important technologies used by our agents are described below: 

\paragraph{Retrieval-Augmented Generation (RAG)}

Enhances generative AI models by integrating external knowledge retrieval \citep{lewis_rag_2021, self_rag}. This approach converts queries into embeddings, matches them with a vectorized knowledge base, and combines retrieved data with generated responses, improving factual accuracy and reducing "hallucinations". Our system employs knowledge graph-based RAGs, which structure data into interconnected graphs \citep{larson_graph_rag_2024}. This method improves the accuracy and relevance of generated content, allowing the model to more efficiently handle data containing a complete description of patterns and the mathematical theory behind the EWP.

\paragraph{Deep Reinforcement Learning (DRL)}

Combines the strengths of both deep learning and reinforcement learning (RL). It has garnered attention for its ability to solve complex problems involving sequential decision-making in high-dimensional spaces . In the traditional RL framework, an agent learns to interact with an environment through a cycle of observing the current state, selecting an action, and receiving feedback in the form of rewards \citep{lapam_drl_2020}. The agent's goal is to learn a policy, which is a strategy for selecting actions that maximizes the cumulative rewards over time. DRL enhances this process by leveraging deep neural networks, a type of machine learning model with multiple layers, to handle and approximate complex functions \citep{kabbani_drl_stock_market_2022, szydlowski_chudziak_hidformer}. Additionally, DRL can address problems with continuous action spaces, where the agent needs to select an action from an infinite set of possibilities, such as adjusting the parameters of a financial trading strategy. 

DRL has been used in the backtesting process to analyze historical market data and learn effective trading strategies \citep{lussange_mas_rl_2020}. By identifying patterns and understanding their impact on future price movements, a DRL agent can make informed decisions to buy, sell, or hold assets, optimizing long-term returns. The ability of DRL to continuously learn and adapt proves particularly valuable in dynamic and uncertain environments, such as financial markets.

\paragraph{Dynamic context}

Refers to the ability of AI agents to adaptively adjust their contextual understanding based on real-time information \citep{dynamic_context}. Agents can utilize various types of context, including tools, documents accessed through RAG, the history of conversations, and the ability to reflect and plan future actions. This approach leverages ongoing interactions and updates the context dynamically, enabling the agent to maintain relevance and accuracy throughout a session. By incorporating new data as it becomes available, dynamic context helps agents refine their responses and improve decision-making processes. 

Memory plays a vital role in enhancing the agent's ability to understand and generate responses based on past interactions, improving decision-making and context-awareness over time. Memory in AI agents, is crucial for handling sequential data and retaining information over long periods \citep{weng_llm_agents_2024}. They achieve this through gated mechanisms that regulate the flow of information, making them highly effective for tasks requiring long-term dependencies, such as time series prediction and natural language processing.

\subsection{Agents Flow Engineering}

Creating effective crews in multi-agent systems involves a combination of strategic orchestration, collaboration, and dynamic task decomposition \citep{guo_llms_mas_survey_2024}. Multiple agents working together to achieve common goal can be managed through orchestration, where a central coordinator assigns tasks and ensures synchronized efforts. This approach enhances control and reliability by allowing the orchestrator to monitor progress, handle exceptions, and optimize resource allocation. In contrast, sequential process allows agents to interact autonomously based on predefined protocols, promoting flexibility and better decision-making.

Agents share information and work collaboratively, either through disordered cooperation (hierarchical process), where agents communicate freely and process inputs in a network-like structure, or through ordered cooperation (sequential process), where agents follow a structured sequence to build on each other’s outputs \citep{li_more_agents_2024}. Our agents operates in a hierarchical mode, allowing for asynchronous execution of tasks. This significantly accelerates the entire prediction process. In this model, higher-level agents decompose complex tasks and delegate subtasks to lower-level agents. Information flows both up and down the hierarchy, with lower-level agents reporting results to their superiors, and higher-level agents providing context and coordination information to their subordinates.

Dynamic scaling is crucial for the adaptability and efficiency of multi-agent systems. By adjusting the number of active agents based on task complexity and available resources, systems can manage workloads more effectively \citep{guo_llms_mas_survey_2024}. Dynamic scaling allows for the autonomous increase or decrease of agents, ensuring optimal resource utilization and maintaining system performance under varying conditions. 

Another part of effective multi-agent system is task decomposition, which enables the breakdown of complex tasks into smaller, manageable sub-tasks. In hierarchical task decomposition organizes tasks into a structured hierarchy, where each level of the hierarchy can be further decomposed until tasks reach a granularity suitable for individual agents \citep{chen_llm_agents_orchestration_2024}. This clarity ensures that agents can focus on their specialized tasks while the orchestrator manages overall coordination. 

\section{Experiments and Results}

\subsection{Data and Use Cases}

The presented system utilizes data from NYSE with a various intervals, allowing for the analysis of the majority of companies listed on this exchange. ElliottAgents provides the flexibility to define the time frame over which the analysis is to be performed, allowing users to conduct both short-term and long-term forecasts.

There are more patterns discovered and described in the Elliott Wave Theory and in our study we have focused on describing only a few selected ones, based on EWP we can distinguish the following use cases:

\begin{itemize}
    \item \textbf{Identifying Impulse Waves}: Impulse waves determine the direction of the main market trend. The hypothesis is that recognizing these impulsive patterns can help predict future price movements. Impulse waves are five-wave patterns that move in the direction of the overall trend, consisting of three actionary waves (1, 3, and 5) and two corrective waves (2 and 4) \citep{frost_ewp_2001}. 
    
    \item \textbf{Identifying ABC Corrections}: In EWP, ABC corrections follows the impulsive move. The hypothesis is that understanding these corrections can provide insights into potential market reversals or continuations. An ABC correction is a three-wave pattern that moves counter to the preceding impulse wave. This pattern helps traders understand when a correction is likely to end and the previous trend will resume.
    
    \item \textbf{Recognize wave extensions}: The objective of this use case is to recognize and analyze wave extensions within Elliott Wave patterns to improve prediction accuracy. Wave extensions, typically seen in the third wave of an impulsive sequence, exceed the standard 1.618 Fibonacci ratio, often reaching up to 2.618 or beyond, indicating a robust trend. 
    
    \item \textbf{Determining support, resistance and target levels}: Support and resistance levels are critical price points where a stock is likely to reverse or pause. Support levels are price points where a downtrend is expected to halt due to a concentration of demand, while resistance levels are where an uptrend is likely to pause due to a concentration of supply. These levels are established by the ending points of previous Elliott waves.
\end{itemize}

\begin{figure}[!t]
    \centering
    \includegraphics[width=\columnwidth]{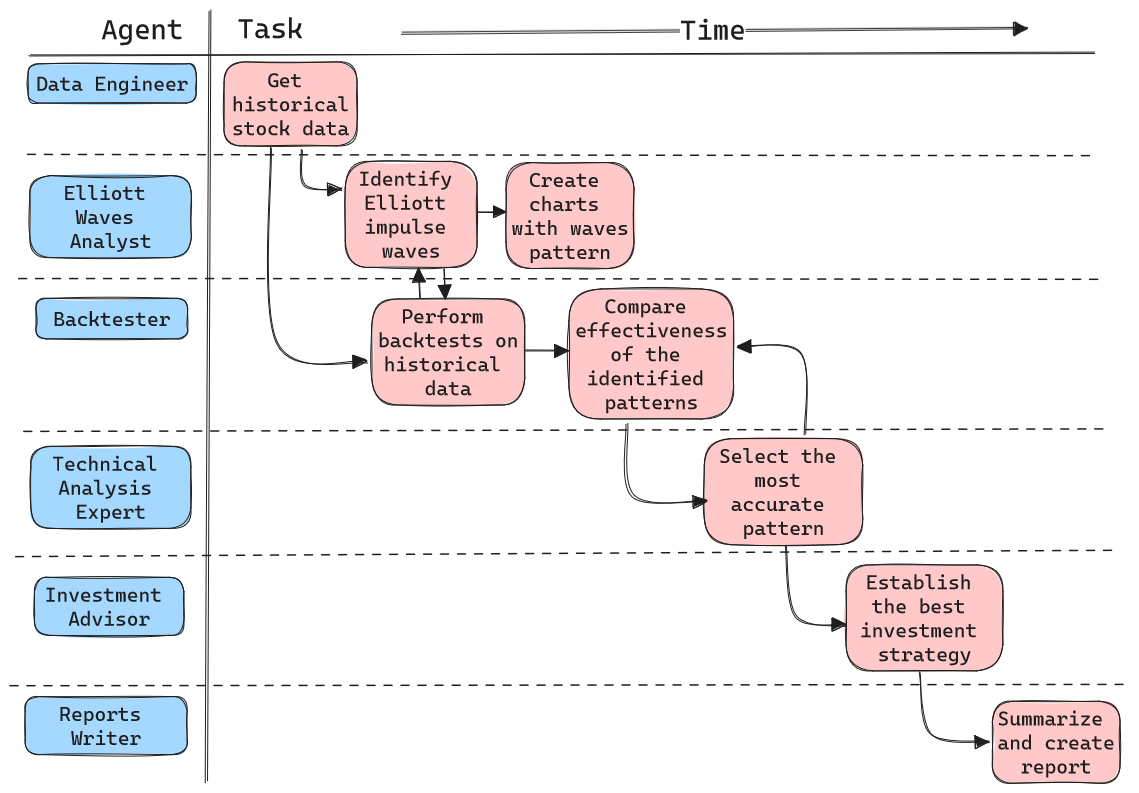}
    \caption{Flow diagram of agents identifying impulse wave use case.}
    \label{fig:use_case_diagram}
\end{figure}

Fig. \ref{fig:use_case_diagram} illustrates a use case diagram showing the workflow of agents in the identification of Elliott impulsive waves. Each agent is assigned specific tasks that are prerequisites for the subsequent agent’s activities, ensuring a seamless and systematic process.

\subsection{Evaluation of Use Cases}

In the first part of experiment, the system tests were conducted using historical data in hourly and daily intervals. The system was run on limited historical data from the largest American companies, with data ranging from one month to two years, to recognize all waves pattern and identify possible buy or sell signals on the charts using knowledge from backtesting process. When the system issued such a signal, we iteratively added additional historical data, allowing the system to detect other patterns and issue another signals. This approach enabled us to evaluate its effectiveness in simulated, but realistic market conditions. Based on these signals, we could simulate transactions and calculate theoretical investment returns, proving the effectiveness of our agent's collaboration.

\begin{figure}[!t]
    \centering
    \includegraphics[width=\columnwidth]{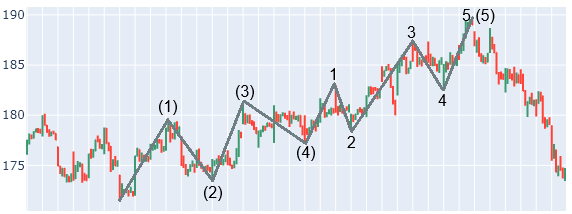}
    \caption{Ending diagonal pattern recognized on AMZN 1h chart.}
    \label{fig:AMZN_Ending_diagonal_1h}
\end{figure}

Fig. \ref{fig:AMZN_Ending_diagonal_1h} presents analysis of Amazon's stock over an approximately two-month period, using hourly intervals. ElliottAgents successfully identified an "ending diagonal" pattern. This pattern, according to EWP, signifies the termination of a larger trend and often precedes a significant reversal in the market direction \citep{frost_ewp_2001}. After confirmation of the trend reversal, ElliottAgents issued a sell recommendation at \$185 per share. The target price was set at \$177 per share, which corresponds to the peak of the second wave extension within the impulsive wave sequence. As illustrated in the accompanying chart, the market behavior adhered closely to our predicted scenario. The theoretical profit from this transaction is \$8 per share, representing a 4.4\% gain, achieved within a short span of just five days. 

\begin{figure}[!t]
    \centering
    \includegraphics[width=\columnwidth]{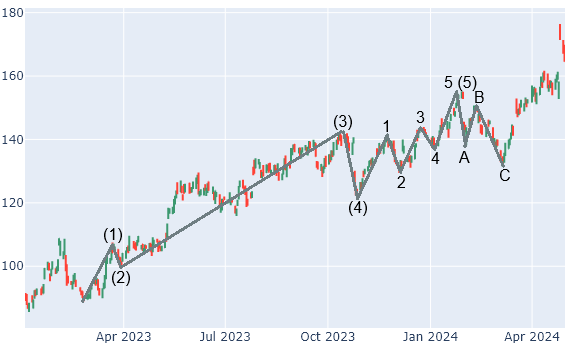}
    \caption{Fifth wave extension found on GOOG 1d chart.}
    \label{fig:GOOG_Fifth_wave_extension_1d}
\end{figure}

Fig. \ref{fig:GOOG_Fifth_wave_extension_1d} presents a results of analysis conducted on Alphabet's stock over a one year period, with data aggregated on a daily interval. ElliottAgents successfully identified multiple patterns during this period. Specifically, the analysis revealed an impulsive wave sequence denoted as (1)-(2)-(3)-(4)-(5), wherein the fifth wave is an extension and a corrective wave pattern, labeled A-B-C immediately after impulsive wave. This corrective pattern terminated at the peak of the second wave of the extension, aligning perfectly with the theoretical expectations posited by EWP \citep{frost_ewp_2001}. According to the theory, the presence of this pattern suggests a forthcoming reversal exceeding the peak of the fifth wave. Upon recognizing this configuration and confirming started reversal, ElliottAgents generated a buy recommendation at a price point of \$140 per share. The target price was strategically set at \$160 per share, aligning with the peak of the fifth wave, while also accounting for the resistance level observed at the peak of wave B (\$150). This dual target strategy ensures both an optimal exit point and a buffer for potential resistance encounters. As the experimental data, in the chart shows, price levels have been achieved. The theoretical profit realized from this transaction amounted to \$20 per share, translating to a 13.3\% gain.

\begin{figure}[!b]
    \centering
    \includegraphics[width=\columnwidth]{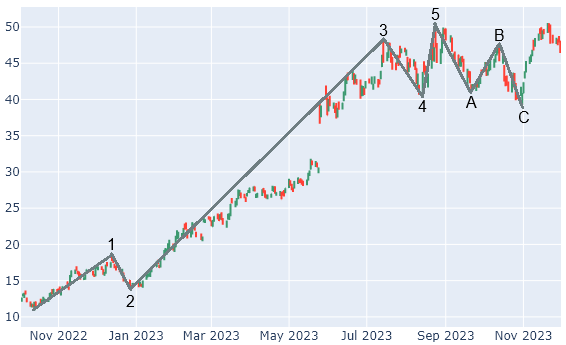}
    \caption{Full wave cycle recognized on NVDA 1d chart.}
    \label{fig:NVDA_Full_cycle_1d}
\end{figure}

In the Fig. \ref{fig:NVDA_Full_cycle_1d} we present a long-term analysis of Nvidia's stock price, on daily intervals. System detected a complete Elliott wave cycle, consisting of an impulsive wave followed by a corrective wave. The identification of this pattern suggests the potential end of the corrective phase and the continuation of the broader trend \citep{murphy_technical_analysis_1999}, which is bullish in this case. The peak of wave C was precisely identified at \$39 per share. Anticipating a reversal in the trend, the system issued a buy recommendation at \$42 per share. The target price was set at \$50 per share, located at the peak of the fifth wave. This target signifies the emergence of the first wave in a new impulsive sequence, according to EWP. The chart clearly demonstrates that a sharp rebound occurred shortly after the recommendation was made, resulting in the target price being reached. This scenario yielded a theoretical profit of \$8 per share, representing a 17.4\% gain.

The second part of the experiment focus on quantity tests for the correctness of the detected pattern and the impact of DRL on results. Test was conducted using a cross-validation method on 1000 samples (candlesticks) with a daily interval for the stocks, which were analyzed in the previous section. In these tests, we focused on examining unfinished impulsive waves (1-2-3-4) consisting of 4 sub-waves as well as complete impulsive waves (1-2-3-4-5), in each case waves could not overlap. In both cases, we compared the results with and without a DRL backtesting process conducted on 10 years of historical data for each company.

\begin{table*}
    \centering
    \begin{tabular}{l | c c c | c c c}
    \hline
    \multirow{2}{*}{\textbf{Stock}} & \multicolumn{3}{c}{\textbf{1-2-3-4 Patterns}} & \multicolumn{3}{| c}{\textbf{1-2-3-4-5 Patterns}} \\
    \cline{2-7}
     & {\bfseries N} & \begin{tabular}[c]{@{}c@{}}{\bfseries Without}\\{\bfseries backtesting}\end{tabular} & \begin{tabular}[c]{@{}c@{}}{\bfseries With}\\{\bfseries backtesting}\end{tabular} & {\bfseries N} & \begin{tabular}[c]{@{}c@{}}{\bfseries Without}\\{\bfseries backtesting}\end{tabular} & \begin{tabular}[c]{@{}c@{}}{\bfseries With}\\{\bfseries backtesting}\end{tabular} \\
    \hline
    \multicolumn{7}{l}{\textit{Daily Interval}} \\
    \hline
    \textbf{AMZN} & 24 & 58.34\% & 66.67\% & 18 & 66.67\% & 77.78\% \\
    \hline
    \textbf{GOOG} & 28 & 53.57\% & 67.86\% & 23 & 65.22\% & 82.61\% \\
    \hline
    \textbf{INTC} & 19 & 57.89\% & 73.68\% & 15 & 60.00\% & 73.34\% \\
    \hline
    \multicolumn{7}{l}{\textit{Hourly Interval}} \\
    \hline
    \textbf{AMZN} & 10 & 50.00\% & 70.00\% & 8 & 62.50\% & 75.00\% \\
    \hline
    \textbf{GOOG} & 13 & 53.84\% & 61.54\% & 9 & 77.78\% & 77.78\% \\
    \hline
    \textbf{INTC} & 12 & 58.34\% & 66.67\% & 9 & 66.67\% & 88.89\% \\
    \hline
    \bottomrule
    \multicolumn{7}{l}{\footnotesize N: number of patterns found.}
    \end{tabular}
    \caption{Comparison of pattern recognition with and without backtesting}
    \label{tab:pattern_comparison}
\end{table*} 

Based on the identified patterns, agents predicted whether the next movement would be upward or downward. A prediction was considered correct if the average price of the subsequent \textit{n} candlesticks was higher or lower, depending on the issued signal. The \textit{n} number of candlesticks was determined according to EWP, where in the case of waves 1-2-3-4, the length of the fifth wave should be approximately 1.62 times the length of the first wave, and in the case of a complete impulse wave, the following wave A should have a length close to wave 5.

Table 1 presents the results of the cross-validation experiments for 1000 data samples in two time intervals. As we can see, the identification of a complete impulsive wave pattern contributes to better predictions of subsequent price movements than incomplete impulse wave pattern. In case of hourly intervals our system detected smaller number of patterns, mainly because price changes on the hourly interval were smaller. The use of DRL resulted in a improvement in prediction, showing that agents are able to use the learning process on historical data in better interpretation of patterns.

\subsection{Success Criteria}

The success criteria for ElliottAgents focus on accurate pattern recognition and analysis to enable users to achieve real market profits. By leveraging NLP, the system can present complex financial information in a clear and understandable manner. The system must provide analysis of a company's stock based on user inputs, identifying all possible wave patterns. These patterns should be visually represented on a chart, and based on them, agents should provide actionable insights, including target price levels based on Fibonacci relationships and key support and resistance levels. Additionally, it should offer clear buy or sell recommendations based on wave analysis, price projections, and trend analysis, with specific time frames for action. Success is measured by the system's ability to deliver pattern recognition and analysis that can be used by traders in investment decisions. 

\section{Discussion and Future Work}

\subsection{Comparison with Other Systems}

Multi-agent architectures have been utilized in stock price prediction systems for many years \citep{akintola_mas_stock_trading_2021, gamil_mas_fuzzy_logic_stock_market_2007, luo_mas_decision_support_stock_market_2002}. However, advancements in AI over recent years have significantly enhanced these systems capabilities. Traditional systems often relied on static rules and fuzzy logic to make decisions, but they faced limitations in accuracy and adaptability. The introduction of fuzzy logic, as seen in older systems, provided a foundation for integrating qualitative judgments with quantitative analysis, yet it required further optimization to improve decision-making. It is difficult to compare the profitability of our system with other price prediction systems available to date. However, based on our experiments, we see that the system can effectively detect and interpret wave patterns, with better accuracy than similar systems using EWP \citep{tirea_mas_ewp_2012}. The analyses created by our agents, clearly present an investment plan, with price levels, that can be used in real world by the traders.

\subsection{Future Enhancements}

Currently, our work has focused primarily on a few patterns recognized by EWP. Expanding our analysis to include additional wave formations such as truncations, zigzags, flat corrections, triangles, and other patterns could significantly enhance our predictive capabilities. Following the successful integration of EWP, we could further improve our system by incorporating other technical analysis methods, such as moving averages. This expansion could enhance our ability to determine more accurate buy or sell signals, potentially improving signal reliability and profitability.

\section{Conclusion}

ElliottAgents demonstrates the potential of integrating NLP and multi-agent systems in the domain of stock market analysis \citep{tunstall_nlp_with_transformers_2022}. By leveraging LLMs and the EWP, the system transforms complex historical market data into comprehensible predictions and explanations. The key innovation lies in the inter-agent dialogue, which mimics collaborative human analysis while harnessing AI's pattern recognition capabilities. This approach not only enhances the accuracy of technical analysis but also addresses the challenge of making financial data interpretable to human users.

Experimental results, conducted on historical data over a period of several years on some of the largest U.S. companies, validate the system's effectiveness in recognizing market patterns and generating natural language descriptions of trends across various time frames. The multi-agent architecture, facilitated by advanced NLP techniques, enables the decomposition of complex analytical tasks, leading to more nuanced and reliable predictions.
This research contributes to the broader field of NLP applications in data-intensive domains, showcasing how AI-driven dialogue systems can enhance collaborative analysis. ElliottAgents bridges the gap between sophisticated AI analysis and human understanding, paving the way for more interpretable and adaptive prediction systems in finance and potentially other specialized fields.

\bibliography{custom}

\end{document}